# Mobility-Aware Content Placement for Device-to-Device Caching Systems

Jaeyoung Song, *Student Member, IEEE,* and Wan Choi, *Senior Member, IEEE*

**Abstract**

User mobility has a large effect on optimal content placement in D2D caching networks. Since a typical user can communicate neighboring users who stay in the D2D communication area of the typical user, the optimal content placement should be changed according to the user mobility. Under consideration of randomness of incoming and outgoing users, we formulate an optimization problem to minimize the average data load of a BS. It is proved that minimization of the average data load of a BS can be transformed to maximization of a monotonic submodular function with a matroid constraint, for which a greedy algorithm can find near-optimal solutions. Moreover, when motions of neighboring users are rapid, the optimal content placement is derived in closed-form, aided by reasonable approximation and relaxation. In the high mobility regime, the optimal content placement is shown to cache partial amounts of the most popular contents.

**Index Terms**

Mobility, Caching, Content placement, Optimal content placement, Wireless edge caching, D2D caching

J. Song and W. Choi are with School of Electrical Engineering, Korea Advanced Institute of Science and Technology (KAIST), Daejeon 34141, Korea (e-mail: wchoi@kaist.edu).

# I. Introduction

As explosive growth of multimedia services such as Netflix and YouTube, the demands for high quality of wireless communication increase tremendously. In fact, more than half of the wireless traffic is video requests, and fortunately, the video traffic is composed of many repeated requests for a few popular contents. Exploiting this feature, wireless caching is touted as a promising technique for future wireless communication systems [2]. In wireless edge caching systems, it is important to determine which contents to cache in which storages. Hence, the recent papers [3]–[7] and references therein investigated optimal content placements. When users have different connectivity to access points (APs) which are equipped with memory, finding an optimal content placement at APs is NP complete [3]. A tradeoff between channel diversity and content diversity was figured out and a greedy algorithm was shown to be optimal for a single user system [4]. For probabilistic content placement, when APs and users are distributed as Poisson point process, an optimal distribution for content caching was studied in [5].

However, the results of the previous literature are basically applicable only to stationary users, not to users with mobility. In fact, users have mobility while video contents are much longer than the time scale of user movement. Hence, users can leave before transmitting or receiving all parts of the requested content. Recently, a few works have investigated the effect of mobility on the content placement. When user mobility is taken into account, [8] explored important factors which affect the performance of caching systems, such as cell sojourn time, contact, and inter-contact time. When the mobility path of a user is given, content placements at small base stations (BSs) were optimized in a distributed manner [9]. Also, in [10], optimal storage allocation was analyzed based on users' cell transitions which was extracted from trajectories of users and modeled as a Markov chain. Treating contact time as a mobility feature, [11] showed that maximizing a data offloading ratio is equivalent to maximizing a submodular function over a matroid constraint. Moreover, when the distribution of contact time is assumed as exponential, the optimal content placement which maximizes a successful delivery probability was investigated in [12]. The authors of [13] conducted a joint optimization of perceived delay and content placement when contact between users follows a Poisson distribution. When contact frequency, the number of contacts in unit time, is given, an offloading probability was maximized in [14].

Based on the randomness of user movement, the contact event is modeled as Poisson event in the aforementioned literature [11]–[14]. Correspondingly, when contact rates of all user pairs are given, the contact time, which accounts for user mobility, is given as an exponential random variable. However, mobility make a distinction from stationarity not only in time domain but also in space domain. As a user moves randomly, the time allowed for communication changes, and at the same time, the location of the user also varies correspondingly.

To accurately capture both space and time effects of user mobility, this paper takes account of the relative movements of a typical user defined as a user of which content placement is optimized and the other users. Given a typical user of interest, other users except for the typical user can be classified into two groups: one is a set of users who moves into the D2D communication range of typical user and, the other is a set of users who leave the range. Consequently, the number of users inside a D2D communication area of the typical user is able to account for the spatial characteristic of mobility. Furthermore, neighboring users who are located in the D2D communication area of the typical user can leave the D2D communication area at any time, so the length of stay is another key parameter that accounts for the time-domain effect of mobility. Considering mobility parameters including the number of users and the length of stay, we investigate an optimal content placement which minimizes average data load of a BS. Minimization of the average data load of a BS is an integer programming problem, generally known to be NP-hard. However, we prove that our minimization problem is equivalent to maximization of a submodular function with a matroid constraint. It is known that a greedy algorithm achieves a constant scale of optimal performance for submodular maximization [15].

This implies that the average data load of a BS can be reduced near-optimally with an appropriate low-complexity algorithm. Furthermore, when neighboring users are moving much faster than the typical user, the optimal content placement is figured out in closed-form aided by continuous relaxation.

In short, our contributions can be summarized as follows.

- Considering user mobility, both temporal and spatial features are accurately modeled and taken into account in wireless edge caching systems. The minimization of the average data load of a BS is formulated as an integer programming problem.
- Minimizing the average data load of a BS is an integer programming problem which is known to be NP-hard. However, we prove that the minimization is equivalent to maximization of a submodular function with a matroid constraint. Also, we propose a low-complexity algorithm which achieves near-optimal performance.
- When neighboring users are moving relatively faster than the typical user, after continuous relaxation, we can find an optimal content placement in closed-form.

The remainder of this paper is organized as follows. The system model is described in Section II. We formulate our problem in Section III. Also, it is shown that our problem is equivalent to maximizing a submodular function under a matroid constraint in Section IV. Section V analyzes optimal content placements in the high mobility regime. Numerical results are provided in Section VI. Finally, we conclude this paper in Section VII.

## II. System Model

In this section, we describe the system model of D2D caching considered in this work.

### *A. D2D Caching*

We consider a D2D communication network in which users can store contents in their memory. Each user requests a content from a library $\mathcal{F} = \{1, 2, \cdots F\}$. The probability of requesting content $i$, $f_i$, is given as $f_i = \frac{i^{-\gamma}}{\sum_{j=1}^{F} j^{-\gamma}}$ for the Zipf exponent $\gamma$. We assume that each content has a unit size and is composed with $L$ packets. Hence, the size of each packet is $\frac{1}{L}$. Also, the memory size of each user is given as $\frac{M}{L}$ so that each user can store up to $M$ packets. In off-peak time when users are idle, packets of contents to be stored are delivered from a BS. Later in non off-peak time, packets stored in users' memory are exploited via D2D communication. Ideally, content placement for all D2D users should be jointly optimized. However, due to users' mobility, the set of users is not fixed and complexity of a joint optimization would be prohibitive. Hence, we consider that each user individually and stochastically optimize the contents to be stored and updates its memory in off-peak time. To this end, we consider content placement of an arbitrary user, who is called a typical user, under assumptions that the number of neighboring users defined as users who are close to the typical user and the contents stored at neighboring users are random. Every user can be a typical user and thus the analytic result for the typical user can be applicable to an arbitrary user in the network in a stochastic sense. The contents cached at a typical user and neighboring users are denoted as $\{c_1^o, c_2^o, \cdots, c_F^o\}$, and $\{d_1^k, d_2^k, \cdots, d_F^k | \ \forall k\}$, in which $c_i^o$ and $d_i^k$ represent the number of packets of the content $i$ that the typical user and the $k$-th neighboring user cache, respectively. Since each content is assumed to be encoded by maximum distance separable (MDS) code so that the packets of each content are MDS coded segments. As a result, any content can be recovered if there exists at least $L$ packets of it [16]. The probability that neighboring user $k$ caches $d_i^k$ packets of content $i$ is defined as $p_{d_i^k}$, and the probability distributions of $p_{d_i^k}$ are generic but independent and identical among neighboring users.

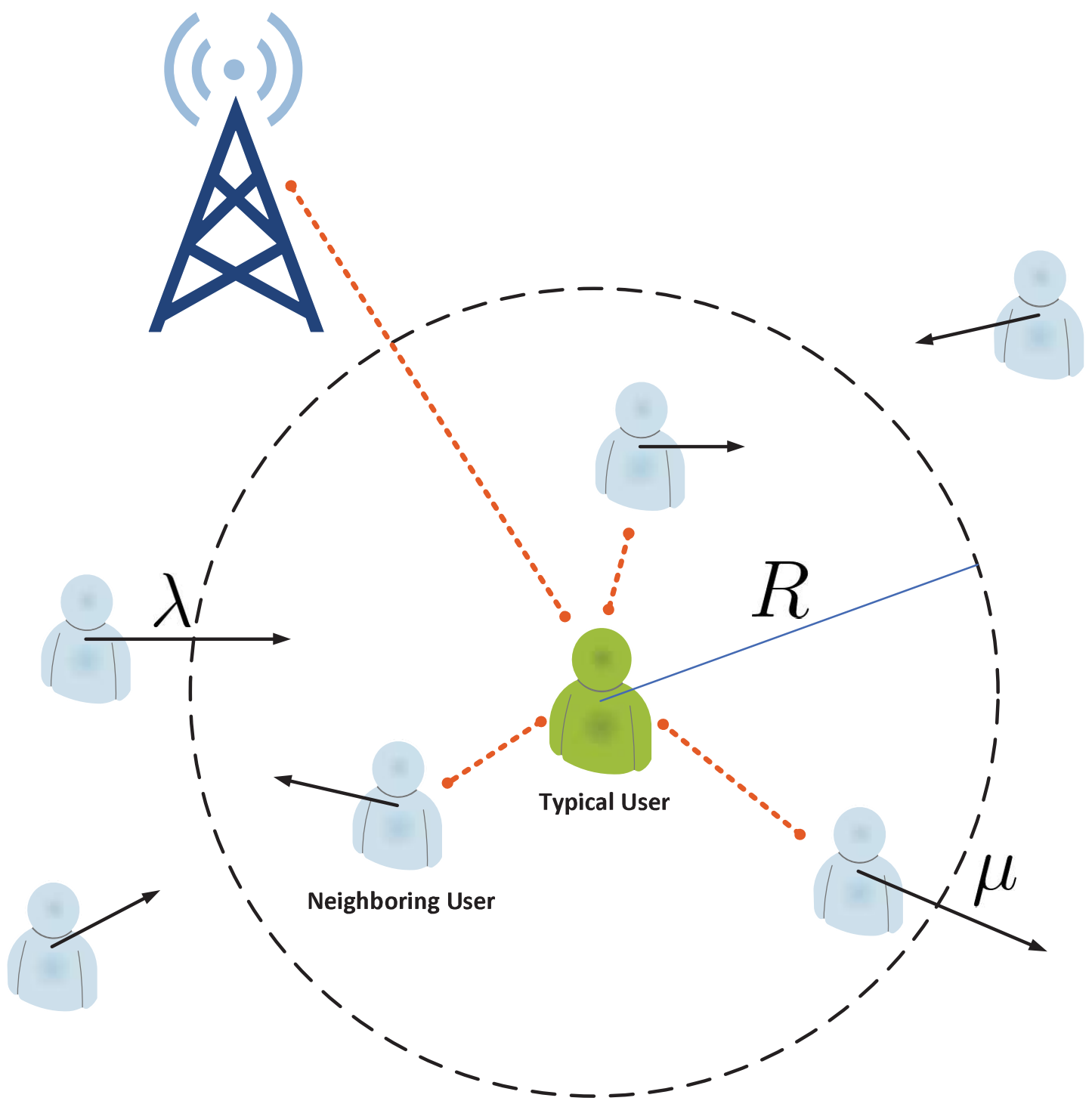

Fig. 1. A mobile D2D caching network. A dashed circle of which radius is $R$ is the area of D2D communication of a typical user. Arrows on each user represent relative velocity with respect to the typical user. $\lambda$ is an incoming rate of move-in users and $\mu$ represents a departure rate of move-out users for the D2D communication area.

### *B. D2D communication*

To communicate via D2D links, proximity between D2D users is required and the range of D2D communication is assumed to be $R$ [12]. Thus, D2D communication is valid inside of a circle with radius $R$ centered at the location of a typical user. When a typical user requests a content in non off-peak time, neighboring users help the typical user by sending the requested content if they cache the content and have enough power to transmit. The other users inside the range of D2D communication of the requesting user cannot request contents until the current request is completely served. Furthermore, neighboring users can transmit to a single user at a time; hence, multicast is not allowed. As a result, the typical user is modeled as a receiver while neighboring users of the typical user are modeled as transmitters. The requests of other neighboring users would be addressed by appropriate scheduling and then another scheduled user becomes a typical user. This paper assume round robin based scheduling so that the effect of scheduling will be the same to all the users, thereby the result for a typical user is valid for any other users.

To satisfy quality of experience (QoE), a fixed rate transmission is adopted. In particular, neighboring users participating D2D communication transmit packets with a rate of $\log(1+\tau)$, where $\tau$ is a SINR threshold for satisfying QoE. Depending on wireless channel fading, distance to the receiver, and other factors, outage can happen; hence, not all transmissions are successful. Furthermore, since there are multiple neighboring users within a range of D2D communication, we consider two different multiple access schemes: orthogonal and non-orthogonal multiple access schemes. For orthogonal multiple access, there exists a central node such as a BS which allocates resource to each of transmitting users in the range of D2D communication orthogonally. Hence, interference can be avoided but resources such as bandwidth and time should be orthogonally divided and used by the number of transmitting users. On the other hand, non-orthogonal multiple access can share entire resource with all the transmitting users at the cost of interference.

In that context, provided that there exist $u$ transmitting users, the average achievable rate $R(u)$ of D2D communication is given as

$$R(u) = \begin{cases} \frac{1}{u}\mathbb{P}\left[\rho > \tau | u = 1\right]\log(1+\tau) & \text{for orthogonal multiple access} \\ \mathbb{P}\left[\rho > \tau | u\right]\log(1+\tau) & \text{for non-orthogonal multiple acces} \end{cases}, \tag{1}$$

where $\rho$ represents SINR as a function of $u$, which is defined as

$$\rho = \frac{HR^{-\alpha}P_t}{I(u-1)+\sigma^2}, \tag{2}$$

where $H$, $R$, $\alpha$, $P_t$, $I(u-1)$, and $\sigma^2$ are wireless fading channel power gain, distance between a transmitting user and the typical user, path-loss exponent, transmit power, interference caused by $u-1$ transmitting users, and noise power.[1] In this work, we assume Rayleigh fading channels, so $H$ is an exponential random variable. Consequently, $\mathbb{P}[\rho > \tau|u]$ is the successful transmission probability for single packet. Also, $u = 1$ implies that a neighboring user is transmitting in given resource block; thus there is no interfering users.

[1]Note that, in this work, neighboring users transmit with fixed power without power control. Successful transmission probability can be enhanced by power control. However, the trend of optimal content placement is not affected because successful transmission probability of all neighboring users increases identically. Hence, for simplicity, power control is not considered.

### C. User mobility

Since users are moving, the number of users able to communicate, $n$, is not fixed. As users' movements are random, there are some users who move into the D2D communication area of the typical user and others who leave the area as shown in Fig. 1. Hence, users entering and leaving the D2D communication area of the typical user can be modeled as arrival and departure processes [17]. We assume arrival and departure processes follow Poisson processes with mean rate $\lambda$ and $\mu$, respectively. However, not all incoming users are available to transmit due to its power constraint. Since mobile devices operate with limited battery power, not all incoming users are capable of transmitting stored packets. In this paper, it is assumed that each user has enough energy to transmit packets with probability $\eta$. Hence, arrival process for users entering the area is thinned by $\eta$. As a consequence, the probability mass function (PMF) of the number of users capable of D2D communication, $p_n$ is given as

$$p_n = \left(\frac{\eta\lambda}{\mu}\right)^n \frac{e^{-\frac{\eta\lambda}{\mu}}}{n!}. \tag{3}$$

Moreover, according to the Poisson departure process with mean rate $\mu$, the expected time of stay in the D2D communication area is given as $T = \frac{1}{\mu}$. Also, users who stay in the D2D communication area are uniformly distributed in the area.

## III. Problem Formulation

Since users are moving and the distance between the transmitting and receiving users is random, not all transmissions of packets are successful. Hence, successful transmission probability for a packet is derived in the following lemma.

***Lemma** 1:* If $u$ transmitting users are uniformly distributed in the D2D communication area, the successful transmission probability for a packet, $\mathbb{P}[\rho > \tau|u]$, is obtained as

$$\mathbb{P}[\rho > \tau|u] = \int_0^R \exp\left(-r^\alpha \left(\frac{P_t}{\sigma^2}\right)^{-1} \tau\right) \left(\int_0^R \frac{x^\alpha}{x^\alpha + \tau r^\alpha} \frac{2x}{R^2} dx\right)^{u-1} \frac{2r}{R^2} dr. \tag{4}$$

*Proof:* Please refer to Appendix A. □

In addition to transmission failures, depending on mobility and content placement of neighboring users, the typical user may not receive the required number of packets of a requesting content. Hence, when content placement of a typical user is optimized in off-peak time in order to reduce the burden of a BS, randomness caused by mobility, wireless channel, and content placement of neighboring users should be taken into account as well as content popularity.

In that context, the data load of a BS can be derived. Define $D(c_i^o, u)$ as the number of packets that a BS needs to serve, provided that $u$ neighboring users are transmitting to the typical user and the typical user requests content $i$. Then, we have

$$D(c_i^o, u) = \left(L - c_i^o - \sum_{k=1}^{u} \min\left\{d_i^k, \left\lfloor \frac{L}{\mu} R(u) \right\rfloor\right\}\right)^+, \tag{5}$$

where $(x)^+ = \max\{0, x\}$. Since the typical user can acquire the packets of content $i$ from own local memory and $u$ neighboring users, a BS delivers the packets that are not supported by D2D or self-caching. When a neighboring user transmits packets, the neighboring user cannot transmit packets not stored in its memory. Also, D2D communication is allowed only when the neighboring user stays in the area. Consequently, the number of packets delivered by D2D can be less than the number of total packets cached in neighboring users' memories or supportable by wireless link quality.

If we average out $D(c_i^o, u)$ over the distributions of typical user's data request, the number of users inside the D2D communication area of the typical user, and the distribution of content placement of neighboring users, the average data load of a BS denoted as $\overline{D(c_i^o)}$ can be obtained as

$$\overline{D(c_i^o)} = \sum_{i=1}^{F} f_i \sum_{n=0}^{\infty} p_n \sum_{\{d_i^1,\cdots,d_i^n\}\in\mathcal{E}} p_{\{d_i^1,\cdots,d_i^n\}} \left(L - c_i^o - \sum_{k=1}^{u} \min\left\{d_i^k, \left\lfloor \frac{L}{\mu} R(u) \right\rfloor\right\}\right)^+, \tag{6}$$

where $p_{\{d_i^1,\cdots,d_i^n\}}$ is the probability that content placements of neighboring users are realized as $\{d_i^1, \cdots, d_i^n\}$ and $\mathcal{E}$ is an event set of all feasible content placements of $n$ neighboring users. Since $\{d_i^k\}$ are independent random variables for different $k$, $p_{\{d_i^1,\cdots,d_i^n\}} = \prod_{k=1}^n p_{d_i^k}$. Also, $u$, which is defined as the number of transmitting users, is equal to the number of non-zero elements of $\{d_i^1, \cdots, d_i^n\}$. (i.e., $u = \|(d_i^1, \cdots, d_i^n)\|_0$.)

Our goal is to find an optimal content placement of the typical user $\{c_i^o\}$ which minimizes the average data load of a BS. Therefore,

$$\mathbf{P1}: \min_{\{c_1^o, c_2^o, \cdots, c_F^o\}} \sum_{i=1}^{F} f_i \sum_{n=0}^{\infty} p_n \sum_{\{d_i^1,\cdots,d_i^n\}\in\mathcal{E}} p_{\{d_i^1,\cdots,d_i^n\}} \left(L - c_i^o - \sum_{k=1}^{u} \min\left\{d_i^k, \left\lfloor \frac{L}{\mu} R(u) \right\rfloor\right\}\right)^+ \tag{7}$$

$$\text{s.t. } c_i^o \in \{0, 1, \cdots L\}, \ \forall i \in \mathcal{F}, \tag{8}$$

$$\sum_{i=1}^{F} c_i^o \leq M. \tag{9}$$

## IV. Submodular Maximization

As our optimization problem **P1** is an integer programming problem which is generally NP-hard, it is hard to solve. However, we can show that **P1** is equivalent to maximization of a submodular function under a matroid constraint. In fact, it is known that a greedy type algorithm achieves at least $1/\epsilon$ of the optimal performance for maximization of a submodular function given constant $\epsilon$ [15]. This means that we can achieve near-optimal performance via a low-complexity algorithm.

In order to prove that **P1** is equivalent to maximization of a submodular function under a matroid constraint, we need to show two statements: the feasible set can be expressed as a matroid, and the objective function is a submodular monotonic function.

First, we show that the constraints of **P1** can be expressed as a matroid. In order to construct a matroid, a ground set is required. Define $S$ as a ground set such that

$$S = \left\{ s_i^l \mid \forall i \in \mathcal{F}, \forall l \in \{1, 2, \cdots, L\} \right\}, \tag{10}$$

where $s_i^l$ represents the $l$-th packet of content $i$. We can divide the ground set $S$ by multiple disjoint sets $S_i$ for $i \in \mathcal{F}$, where $S_i$ is defined as $S_i = \left\{ s_i^l \mid \forall l \in \{1, 2, \cdots, L\} \right\}$. Then, any content placement $\left\{ c_i^o \mid \forall i \right\}$ can be expressed as a subset of $S$. Thus, we define a content placement $C \subseteq S$. Correspondingly, we can also partition the content placement for each content by defining $C_i$ as $C_i = C \cap S_i$, where $C_i$ specifies the indexes of packets of file $i$ to be stored. Additionally, an independent set $\mathcal{I}$ can be defined as

$$\mathcal{I} = \{ C \subseteq S \mid |C| \leq M \}. \tag{11}$$

Note that any feasible content placement is an element of an independent set $\mathcal{I}$. With the following proposition, we show that $(S, \mathcal{I})$ forms a matroid $\mathcal{M}$.

*Proposition 1:* The constraints of **P1**, (8) and (9), can be expressed as matroid $\mathcal{M} = (S, \mathcal{I})$ with the ground set $S$ in (10) and the independent set $\mathcal{I}$ in (11).

*Proof:* Please refer to Appendix B. □

With Proposition 1, the following problem **P1′** is equivalent to **P1** after converting minimization into maximization by inserting negative sign on the objective function.

$$\mathbf{P1'}: \max_{\{c_i^o\}} - \sum_{i=1}^{F} f_i \sum_{n=0}^{\infty} p_n \sum_{\{d_i^1, \cdots, d_i^n\} \in \mathcal{E}} p_{\{d_i^1, \cdots, d_i^n\}} D(C_i, u) \tag{12}$$

$$\text{s.t. } C \in I, \tag{13}$$

where $C_i = C \bigcap S_i$.

Next, we show that the objective function of **P1′** in (12) is a submodular monotonic function. Monotonicity is trivial because if we add an additional packet of any contents to memory of the

**Algorithm 1** The Greedy Algorithm

1: **initialize** $C = \phi$

2: **for** $m = 1, 2, \cdots, M$ **do**

3: $\quad s_k^t = \text{argmax}_{s_i^l \in S} \sum_{i=1}^{F} f_i \sum_{n=0}^{\infty} p_n \sum_{\{d_i^1,\cdots,d_i^n\}\in\mathcal{E}} p_{\{d_i^1,\cdots,d_i^n\}} \left(D(C,u) - D(C \cup \{s_i^l\}, u)\right)$

4: $\quad C \leftarrow C \cup \{s_k^t\}$

5: **end for**

typical user, a data load of a BS is always decreasing. Hence, we prove only the submodularity of the average data load of a BS with Proposition 2.

*Proposition 2:* The objective function of $\mathbf{P1'}$, (12), is a submodular function.

*Proof:* Please refer to Appendix C. □

Based on Propositions 1 and 2, we show that $\mathbf{P1'}$ is maximization of submodular monotonic function under a matroid constraint. Consequently, we can obtain near-optimal performance with a greedy algorithm presented in Algorithm 1. Algorithm 1 finds $M$ packets which maximizes the marginal gain in each step. Hence, the complexity of the Greedy algorithm is given as $O(M)$.

## V. High Mobility Regime

Utilizing the submodular property, we can find a near-optimal solution for minimization of the average data load of a BS. In this section, the structure of optimal solution is analyzed in detail in the high mobility regime. If neighboring users are moving much faster than the typical user, neighboring users leave the D2D communication area rapidly. Hence, the expected time of stay becomes much shorter as the mobility of neighboring users increases. Consequently, for the high mobility regime, we can assume that communication duration is limited. $T \ll 1$. Since the communication duration is not long enough, neighboring users cannot deliver all the packets stored; but only the number of packets that time allows to communicate can be transferred to the typical user. In other words, in the high mobility regime, the bottleneck of the minimum in

(6) becomes the number of packets that can be delivered via the wireless D2D communication, regardless of how many packets are stored in neighboring users, if at least a packet of the requested content is stored. Consequently, for $T \ll 1$ and $d_i^k > 0$,

$$\min\left\{d_i^k, \left\lfloor \frac{L}{\mu} R(u) \right\rfloor\right\} = \left\lfloor \frac{L}{\mu} R(u) \right\rfloor. \tag{14}$$

Using (14), the average data load of a BS, (7), can be reduced to

$$\overline{D(c_i^o)} = \sum_{i=1}^{F} f_i \sum_{n=0}^{\infty} p_n \sum_{\{d_i^1,\cdots,d_i^n\}\in\mathcal{E}} p_{\{d_i^1,\cdots,d_i^n\}} \left(L - c_i^o - u \left\lfloor \frac{L}{\mu} R(u) \right\rfloor\right)^+. \tag{15}$$

For given $u$ transmitting users, the average rate of D2D communication $R(u)$is different for the types of multiple access schemes. Accordingly, analysis is conducted for each multiple access scheme.

*1) Orthogonal Multiple Access:* For orthogonal multiple access, $R(u)$ becomes

$$R(u) = \frac{1}{u} \mathbb{P}\left[\rho > \tau | u = 1\right] \log(1 + \tau). \tag{16}$$

Thus, the average data load of a BS can be written as

$$\overline{D(c_i^o)} = \sum_{i=1}^{F} f_i \sum_{n=0}^{\infty} p_n \sum_{\{d_i^1,\cdots,d_i^n\}\in\mathcal{E}} p_{\{d_i^1,\cdots,d_i^n\}} \left(L - c_i^o - u \left\lfloor \frac{L}{u\mu} \mathbb{P}\left[\rho > \tau | u = 1\right] \log(1 + \tau) \right\rfloor\right)^+. \tag{17}$$

The following lemma can be used to derive the bound of the average data load of a BS for orthogonal multiple access.

***Lemma** 2:* For $g(u) = L - c_i^o - u \left\lfloor \frac{L}{u\mu} \mathbb{P}\left[\rho > \tau | u = 1\right] \log(1 + \tau) \right\rfloor$, $\exists c$ such that $\forall u$, $g(u) \geq T \cdot c$, where $T$ is the time of stay in the D2D communication area.

*Proof:* Please refer to Appendix D. □

Based on Lemma 2, we have the following proposition.

*Proposition 3:* The gap between the average data load of a BS and its lower bound obtained from Jensen's inequality is bounded by the length of stay $T$. In other words,

$$\left| \overline{D(c_i^o)} - \sum_{i=1}^{F} f_i \left( L - c_i^o - \sum_{n=0}^{\infty} p_n \sum_{\{d_i^1,\cdots,d_i^n\}\in\mathcal{E}} p_{\{d_i^1,\cdots,d_i^n\}} u \left\lfloor \frac{L}{u\mu} \mathbb{P}\left[\rho > \tau \mid u = 1\right] \log\left(1 + \tau\right) \right\rfloor \right)^+ \right| < T \cdot |c| \tag{18}$$

*Proof:* Please refer to Appendix E. □

Consequently, in the high mobility regime, where $T$ is small, the gap between the approximated and original objectives approaches to zero. This validates the approximation of the average data load of a BS with the lower bound obtained from Jensen's inequality. For tractability, discrete variables $\{c_i^o\}$ are relaxed to be continuous variables. Hence, the minimization of the average data load of a BS is relaxed to the following problem.

$$\mathbf{P2}: \min_{c_i^o} \sum_{i=1}^{F} f_i \cdot \left( L - c_i^o - \sum_{n=0}^{\infty} p_n \sum_{\{d_i^1,\cdots,d_i^n\}\in\mathcal{E}} p_{\{d_i^1,\cdots,d_i^n\}} u \left\lfloor \frac{L}{u\mu} \mathbb{P}\left[\rho > \tau | u = 1\right] \log(1+\tau) \right\rfloor \right)^{+} \tag{19}$$

$$\text{s.t. } 0 \le c_i^o \le L, \ \forall i \in \mathcal{F}, \tag{20}$$

$$\sum_{i=1}^{F} c_i^o \le M. \tag{21}$$

Now, we can find a property of optimal content placement. Leveraging the property, we can figure out the optimal solution in closed-form.

*Proposition 4:* The optimal solution of **P2**, $\left\{c_{i,\text{opt}}^o\right\}$, has to satisfy the following inequality

$$c_{i,\text{opt}}^o \le L - \nu, \quad \forall i \in \mathcal{F}. \tag{22}$$

where $\nu$ is defined as

$$\nu = \sum_{n=0}^{\infty} p_n \sum_{\{d_i^1,\cdots,d_i^n\}\in\mathcal{E}} p_{\{d_i^1,\cdots,d_i^n\}} u \left\lfloor \frac{L}{u\mu} \mathbb{P}\left[\rho > \tau | u = 1\right] \log(1+\tau) \right\rfloor \tag{23}$$

*Proof:* Please refer to Appendix F. □

Proposition 4 implies that caching a content more than a certain threshold is a waste. Intuitively, the average data load of a BS consists of a summation of the positive values of functions. In detail, $(x)^+$ is always 0 if $x$ is less than 0. Therefore, caching contents more than the threshold does not reduce the average data load of a BS more. Using the aforementioned proposition, we can find the optimal solution.

*Theorem 1:* The optimal solution of **P2**, $\left\{c_{i,\mathrm{OM}}^{o}\right\}$, is obtained as

$$c_{i,\mathrm{OM}}^{o} = \begin{cases} L-\nu & \text{for } \ i<k \\ M-(k-1)(L-\nu) & \text{for } \ i=k \\ 0 & \text{for } \ i>k \end{cases}, \tag{24}$$

where $k = \left\lceil \frac{M}{L-\nu} \right\rceil$.

*Proof:* Please refer to Appendix G. □

In the high-mobility regime, the bottleneck of being served by neighboring users comes from wireless communication aspects. In that context, Theorem 1 proves that the optimal content placement is similar to caching the most popular contents. However, the optimal content placement $\left\{c_{i,\mathrm{OM}}^{o}\right\}$ does not allow to cache a whole content. Since the typical user can receive packets from neighboring users via D2D communication, only the amount of contents that neighboring users cannot support is cached at the typical user.

*2) Non-orthogonal Multiple Access:* When a non-orthogonal multiple access scheme is used, $R(u)$ for non-orthogonal multiple access is given as

$$R(u) = \mathbb{P}\left[\rho > \tau | u\right] \log(1+\tau). \tag{25}$$

Correspondingly, the average data load of a BS is written as

$$\overline{D(c_i^o)} = \sum_{i=1}^{F} f_i \sum_{n=0}^{\infty} p_n \sum_{\left\{d_i^1,\cdots,d_i^n\right\}\in\mathcal{E}} p_{\left\{d_i^1,\cdots,d_i^n\right\}} \left(L - c_i^o - u\left\lfloor \frac{L}{\mu}\mathbb{P}\left[\rho>\tau|u\right]\log(1+\tau)\right\rfloor\right)^{+}. \tag{26}$$

Similar to the orthogonal multiple access chase, we have the following lemma for non-orthogonal multiple access.

***Lemma** 3:* For $g'(u) = L - c_i^o - u\left\lfloor \frac{L}{\mu}\mathbb{P}\left[\rho>\tau|u\right]\log(1+\tau)\right\rfloor$, $\exists c'$ such that $\forall u$, $g'(u) \geq T\cdot c'$, where $T$ is the time of stay in the D2D communication area.

*Proof:* Please refer to Appendix H. □

Lemma 3 implies that $g'(u)$ is bounded below. Hence, we have the following proposition.

*Proposition 5:* The gap between $\overline{D(c_i^o)}$ and its lower bound obtained from Jensen's inequality is bounded by the length of stay $T$. In other words,

$$\left|\overline{D(c_i^o)} - \sum_{i=1}^{F} f_i \left( \sum_{n=0}^{\infty} p_n \sum_{\{d_i^1,\cdots,d_i^n\}\in\mathcal{E}} p_{\{d_i^1,\cdots,d_i^n\}} g'(u) \right)^+ \right| \leq T \cdot |c'|, \tag{27}$$

where $c'$ is the constant given in Lemma 3.

*Proof:* Replacing $g(u)$ with $g'(u)$, we follow the proof of Proposition 3. □

In the high mobility regime, similar to orthogonal multiple access case, the lower bound obtained from Jensen's inequality can be a reasonable approximation of the average data load of a BS. Furthermore, after relaxation of the integer constraint, we have the following problem.

$$\mathbf{P3}: \min_{c_i^o} \sum_{i=1}^{F} f_i \left( L - c_i^o - \frac{L}{\mu} \log(1+\tau) \sum_{n=0}^{\infty} p_n \sum_{\{d_i^1,\cdots,d_i^n\}\in\mathcal{E}} p_{\{d_i^1,\cdots,d_i^n\}} u \mathbb{P}\left[\rho > \tau | u\right] \right)^+ . \tag{28}$$

$$\text{s.t. } 0 \leq c_i^o \leq L, \ \forall i \in \mathcal{F}, \tag{29}$$

$$\sum_{i=1}^{F} c_i^o \leq M. \tag{30}$$

However, **P3** is exactly the same as **P2** if we replace $\frac{L}{\mu}\mathbb{P}\left[\rho > \tau | u = 1\right] \log(1+\tau)$ with $\frac{L}{\mu}\log(1+\tau) \sum_{n=0}^{\infty} p_n \sum_{\{d_i^1,\cdots,d_i^n\}\in\mathcal{E}} p_{\{d_i^1,\cdots,d_i^n\}} u \mathbb{P}\left[\rho > \tau | u\right]$. Hence, the optimal content placement for non-orthogonal multiple access in the high mobility regime can be found in the following theorem.

*Theorem 2:* The optimal solution of **P2**, $\left\{c_{i,\mathrm{NOM}}^o\right\}$, is given as

$$c_{i,\mathrm{NOM}}^o = \begin{cases} L - \zeta & \text{for } i < k \\ M - (k-1)(L-\zeta) & \text{for } i = k \\ 0 & \text{for } i > k \end{cases}, \tag{31}$$

where $\zeta$ is defined as

$$\zeta = \frac{L}{\mu} \log(1+\tau) \sum_{n=0}^{\infty} p_n \sum_{\{d_i^1,\cdots,d_i^n\}\in\mathcal{E}} p_{\{d_i^1,\cdots,d_i^n\}} u \mathbb{P}\left[\rho > \tau | u\right]. \tag{32}$$

*Proof:* Replacing $\nu$ with $\zeta$, we follow the proof of Theorem 1. □

Similar to the orthogonal multiple access case, when neighboring users are moving fast, communication duration imposes restrictions on obtaining the content requested by the typical user. Consequently, contents that neighboring users store cannot be delivered completely. Therefore, the structure of the optimal content placement is not changed in non-orthogonal multiple access. However, since the amount of data that D2D communications support can be different depending on multiple access schemes, the optimal content placement for non-orthogonal multiple access shows different margins from the case of orthogonal multiple access. For the low and intermediate mobility regimes, optimal content placement is highly dependent on the distribution of neighboring users' content placements. Therefore, Algorithm 1 can be used to find the optimal content placement.

## VI. Numerical Results

In this section, we demonstrate performance of the greedy algorithm. Also, the average data load of a BS is quantified via simulations. Depending on environments, our greedy algorithm can achieve near-optimal performance. As the mobilty of neighboring users increases, the performance gap between the optimal solution and the approximated solution in the high mobility regime reduces rapidly.

Default environments for simulations are set as $F = 5$, $\gamma = 0.6$, $M = 5$, $L = 5$, $\eta = 0.5$, $\tau = 5$ [dB], $R = 5$, and $\alpha = 4$. Also, the distributions of neighboring users' content placements are assumed to be uniform for each content (i.e., $p_{d_i^k} = \frac{1}{L+1}, \forall i, k$). Depending on other parameters, the simulation environments are slightly changed and mentioned. correspondingly.

The normalized average data load of a BS as a function of transmit SNR of neighboring users is shown in Fig. 2 when $\lambda = 1$ and $\mu = 1$ and in Fig. 3 when $\lambda = 2$, and $\mu = 2$, respectively.

It is verified that the performance by the greedy algorithm is close to the optimal one obtained via exhaustive search. This implies that the proposed algorithm can be a practical substitution for exhaustive search. Fig. 2 also exhibits that non-orthogonal multiple access is better than orthogonal multiple access in the low and intermediate SNR regimes. For the low and intermediate SNR, non-orthogonal multiple access reduces the average data load of a BS more due to its high efficiency of resource utilization. However, in high SNR, increasing transmit power causes higher interference to the typical user. Hence, avoiding interference via orthogonal allocation of resource is more reliable than utilizing larger bandwidth in high SNR, especially when the number of incoming users are large as shown in Fig 3.

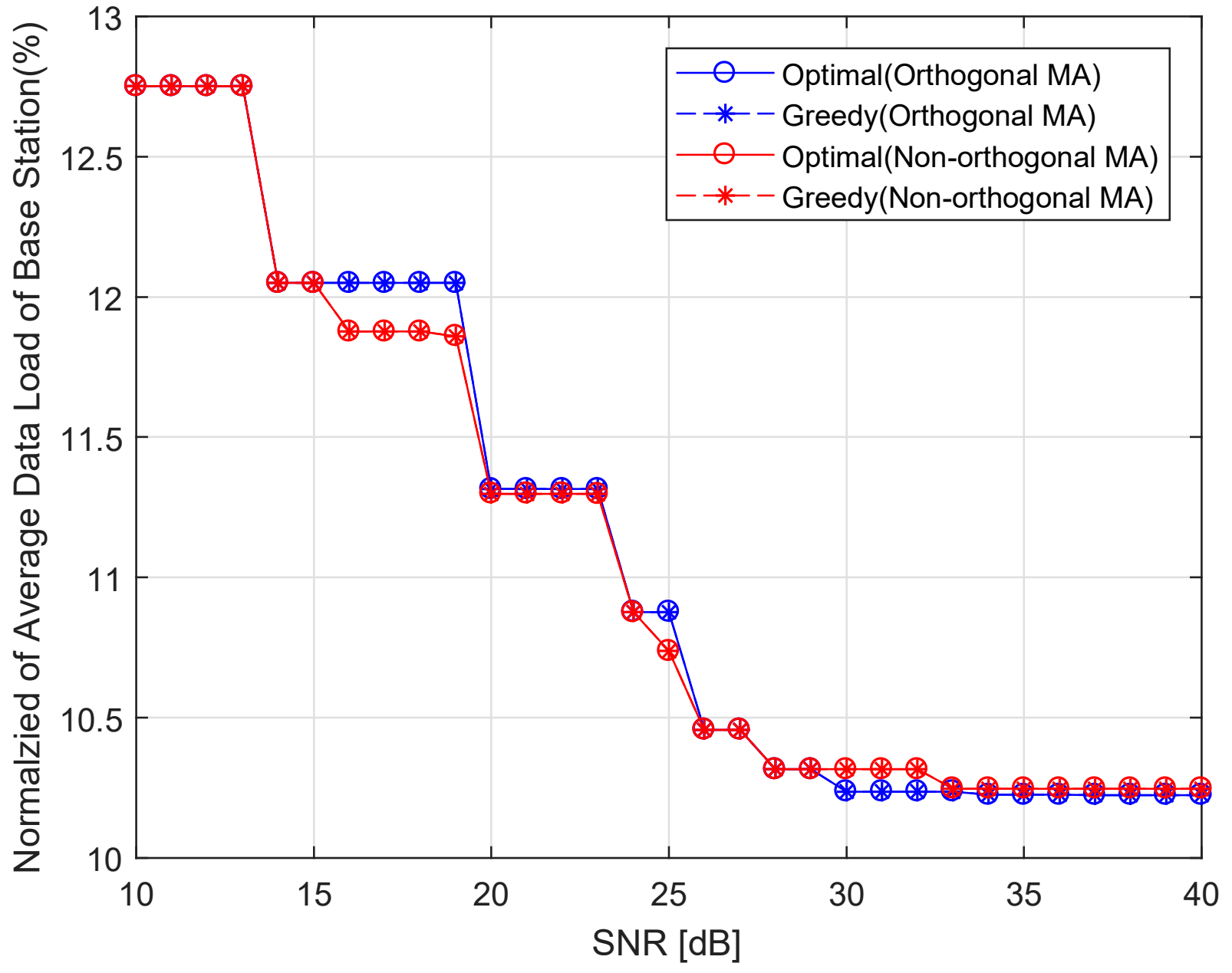

Fig. 2. The normalized average data load of a BS versus SNR for $\lambda = 1$ and $\mu = 1$

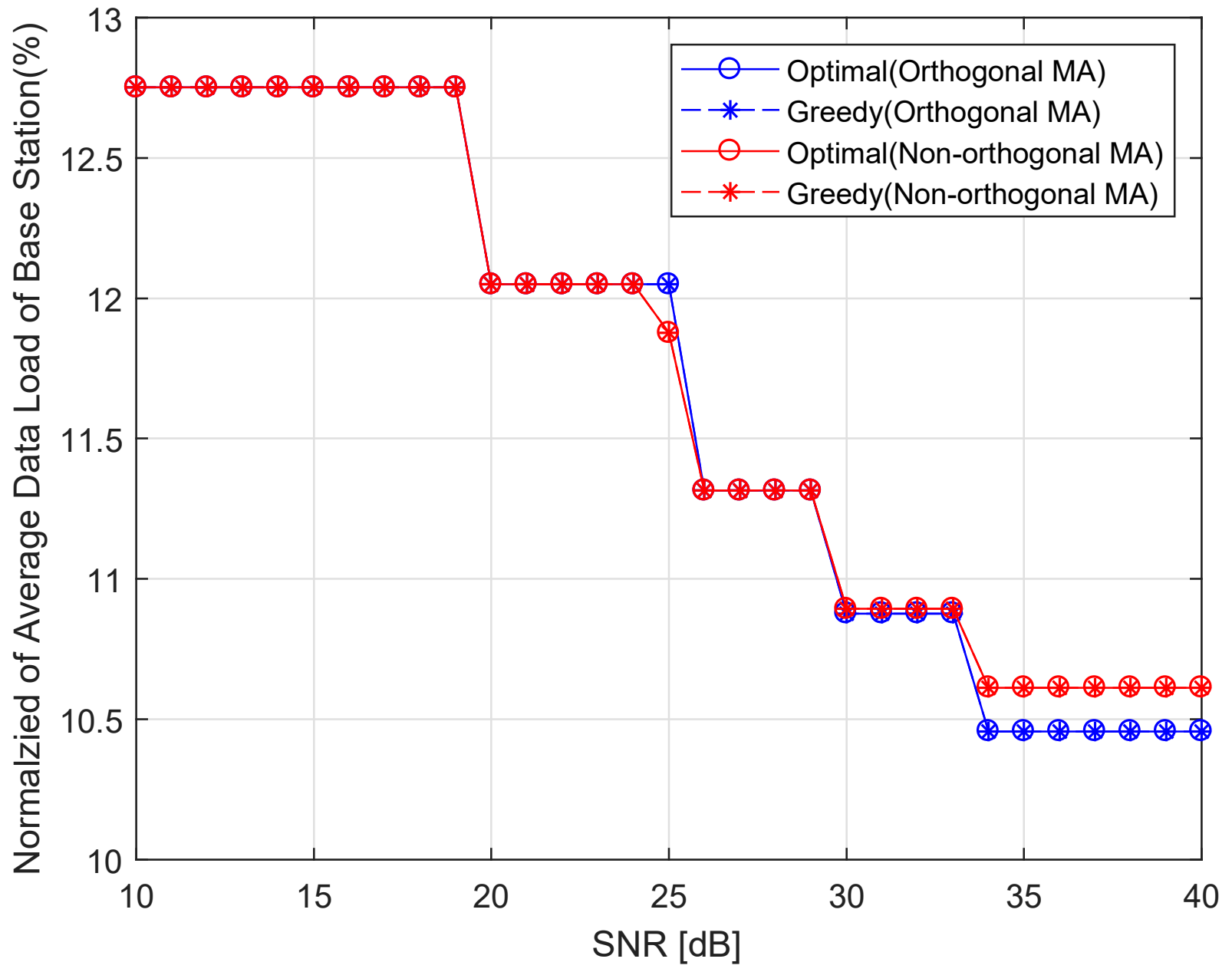

Fig. 3. The normalized average data load of a BS versus SNR for $\lambda = 2$ and $\mu = 2$

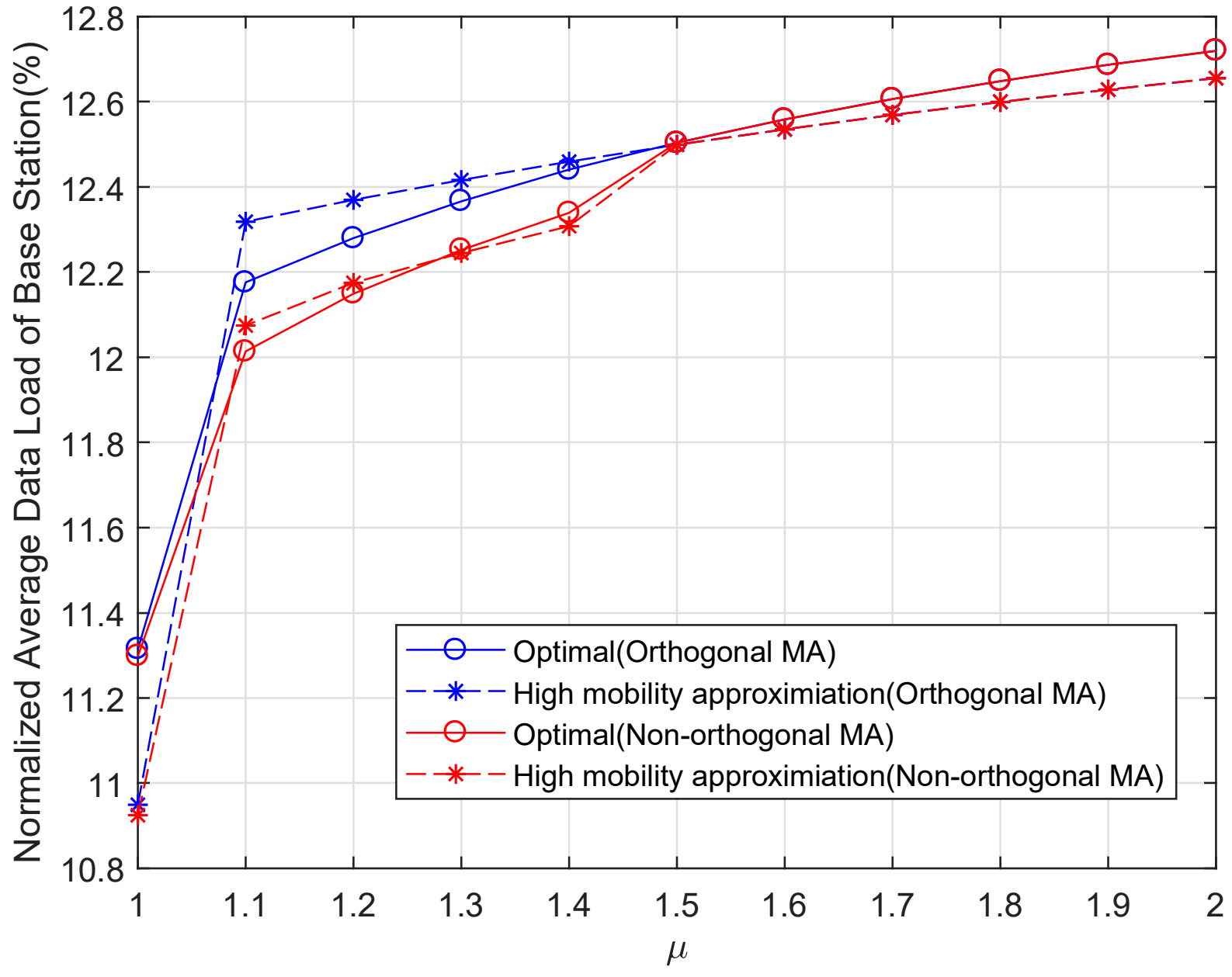

Fig. 4. The normalized average data load of a BS versus mean departure rate for $\mu = 1$ and SNR = 20 [dB]

The average data load of a BS for varying $\mu$ is shown in Figs. 4 and 5 for SNR = 20 [dB] and for SNR = 30 [dB], respectively. The incoming rate is set to be $\lambda = 1$. The mean departure rate $\mu$ is the inverse of the average length of stay of neighboring users in the D2D communication area of the typical user. As $\mu$ increases, the gap between the optimal solution and the high-mobility approximation reduces rapidly. Moreover, when $\mu$ is small, the average data load of a BS increases rapidly because the amount of packets supported via D2D communication reduces fast. However, the average data load of a BS is bounded above; thus it becomes saturated for larger $\mu$. Compared with Fig. 4, Fig. 5 shows that the gap reduces slowly. Since the impact of staying time is less dominant as transmit SNR increases, the gap is slowly decreasing as $\mu$ increases.

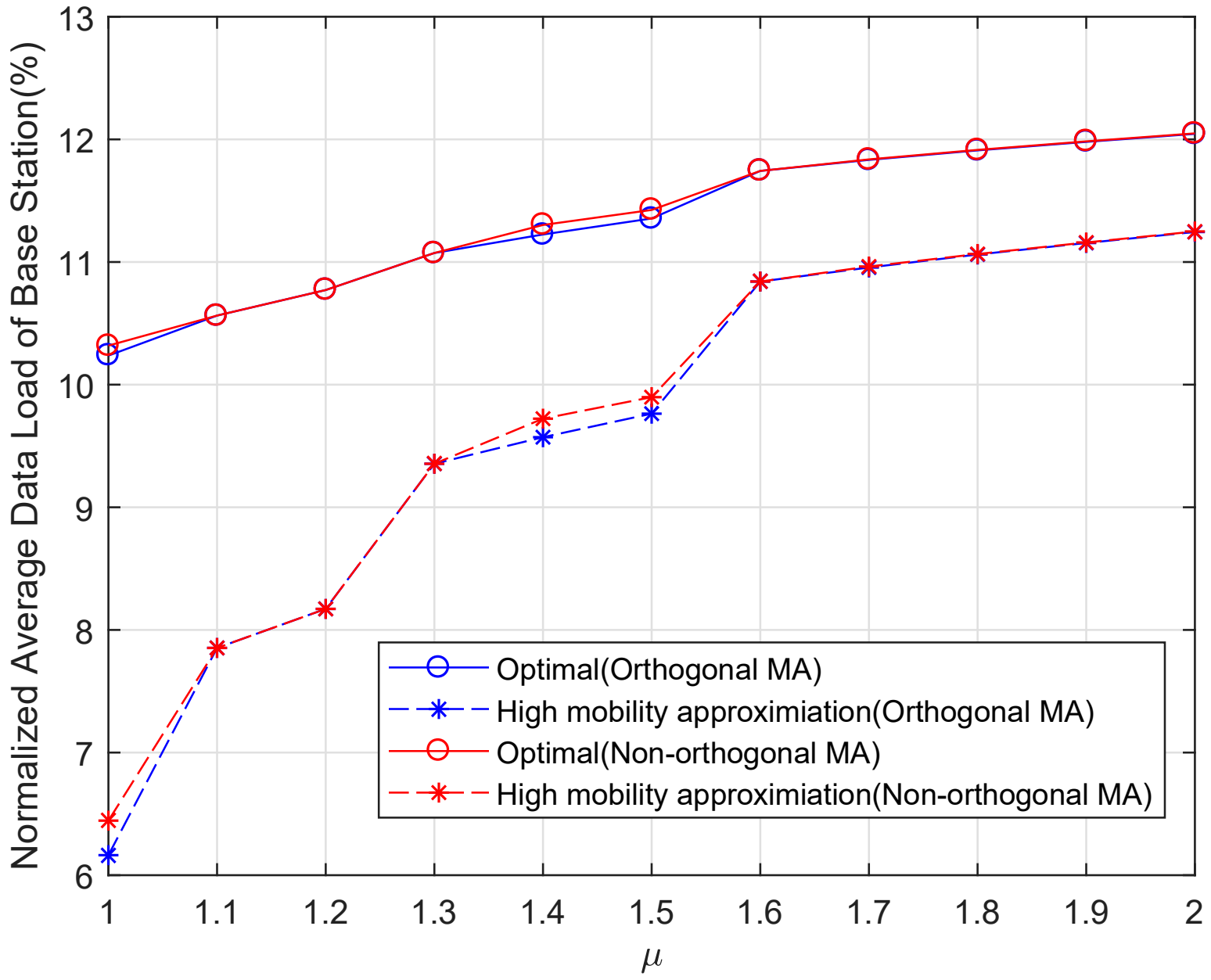

Fig. 5. The average data load of a BS versus mean departure rate for $\mu = 1$ and SNR = 40 [dB]

## VII. Conclusion

For a given typical user and D2D communication area of the typical user, motions of other users were modeled as incoming or outgoing to/from the D2D communication area. With the developed model, we have established the minimization of the average data load of a BS in D2D caching enabled networks. Also, we have shown that the optimization can be transformed to the maximization of a submodular function with a matroid constraint. Owing to the transformation, we could use a greedy algorithm to achieve near-optimal performance. Furthermore, when neighboring users mobility is much higher than the typical user, we have figured out that the optimal content placement is to cache partial amounts of the most popular contents.

# Appendix A

## Proof of the Lemma 1

Using Bayes' rule and the total probability theorem, the successful transmission probability can be expanded as

$$\mathbb{P}\left[\rho > \tau \mid u\right] = \mathbb{E}_R \mathbb{E}_I \left[\mathbb{P}\left[H > R^{\alpha} P_t^{-1} \tau \left(I + \sigma^2\right) |R, I, u\right]\right]. \tag{A.1}$$

Since we assume that $H$ is exponentially distributed,

$$\mathbb{P}\left[\rho > \tau \mid u\right] = \mathbb{E}_R \left[\exp\left(-R^{\alpha} \left(\frac{P_t}{\sigma^2}\right)^{-1} \tau\right) \mathcal{L}_I \left(R^{\alpha} P_t^{-1} \tau\right)\right]. \tag{A.2}$$

Given $u$ transmitting users, there exist $u-1$ users who cause interference. Hence, the Laplace transform of interference is given as

$$\mathcal{L}_I(s) = \mathbb{E}_I\left[e^{-sI}\right], \tag{A.3}$$

$$= \mathbb{E}_{H_i, R_i}\left[e^{-s \sum_{i=1}^{u-1} H_i R_i^{-\alpha} P_t}\right]. \tag{A.4}$$

where $H_i$ and $R_i$ are the Rayleigh fading channel power gain and the distance to the typical user of the $i$-th neighboring user, respectively. Since $u$ neighboring users are uniformly distributed in the D2D communication area, we can rewrite the Laplace transform of interference as

$$\mathcal{L}_I(s) = \left(\mathbb{E}_{R_i}\left[\frac{1}{1 + sR_i^{-\alpha} P_t}\right]\right)^{u-1}, \tag{A.5}$$

$$= \left(\int_0^R \frac{1}{1 + sx^{-\alpha} P_t} \frac{2x}{R^2} dx\right)^{u-1}. \tag{A.6}$$

As a result, the successful transmission probability given $u$ transmitting users becomes

$$\mathbb{P}\left[\rho > \tau \mid u\right] = \mathbb{E}_R \left[\exp\left(-R^{\alpha} \left(\frac{P_t}{\sigma^2}\right)^{-1} \tau\right) \mathcal{L}_I \left(R^{\alpha} p^{-1} \tau\right)\right], \tag{A.7}$$

$$= \int_0^R \exp\left(-r^{\alpha} \left(\frac{P_t}{\sigma^2}\right)^{-1} \tau\right) \left(\int_0^R \frac{x^{\alpha}}{x^{\alpha} + \tau r^{\alpha}} \frac{2x}{R^2} dx\right)^{u-1} \frac{2r}{R^2} dr. \tag{A.8}$$

## Appendix B

## Proof of the Proposition 1

Since any content placement can be expressed as a collection of packets, each subset of $S$ can be a content placement. Since MDS coded packets are cached, a content placement which has $l$ packets of content $i$ can be expressed as a set of packets up-to the $l$-th packet of content $i$. Hence, $\{c_i^o|\forall i\}$ is equivalent to $C = \left\{s_i^m|1 \le m \le c_i^o, \forall i\right\}$. Also, given $C$, $c_i^o$ is equal to $|C \cap S_i|$. (i.e., $c_i^o = |C_i|$.) Therefore, for arbitrary content placement $\left\{c_i^o|\ \forall i\right\}$, we can find equivalent $C$. Moreover, the total number of packets for content placement $C$ can be represented as the cardinality of $C$. Hence, the constraint for memory size (9) can be represented as $|C| \le M$.

Consequently, any feasible content placement is an element of an independent set $\mathcal{I}$ in (11). Also, it can be easily proven that $\mathcal{I}$ satisfies the following conditions for a matroid.

1) $\mathcal{I}$ is nonempty.
2) If $Y \in \mathcal{I}$ and $X \subseteq Y$ then, $X \in \mathcal{I}$.
3) If $X, Y \in \mathcal{I}$, and $|X| < |Y|$ then, $\exists y \in Y \setminus X$ such that $X \cup \{y\} \in \mathcal{I}$.

As a result, (8) and (9) can be expressed as a matroid $\mathcal{M} = (S, \mathcal{I})$.

## Appendix C

## Proof of the Proposition 2

Since the sum of submodular functions is a submodular function, the objective of **P1′**, (12), becomes the submodular function if we prove the following, for $C' \subseteq C$ and $\forall s_i^l \in \mathcal{S} \setminus C$,

$$G\left(C_i \cup \left\{s_i^l\right\}\right) - G\left(C_i\right) \le G\left(C_i' \cup \left\{s_i^l\right\}\right) - G\left(C_i'\right), \tag{C.1}$$

where $G(C_i) = -f_i p_n p_{\{d_i^1,\cdots,d_i^n\}} D(C_i, u)$.

To show (C.1), we rewrite $G(C_i)$ as

$$G(C_i) = -\beta\left(Q(C_i, u)\right)^+, \tag{C.2}$$

where $\beta = f_i p_n p_{\{d_i^1,\cdots,d_i^n\}}$ and $Q(C_i, u) = L - |C_i| - \sum_{k=1}^{u} \min\left\{d_i^k, \left\lfloor \frac{L}{\mu} R(u) \right\rfloor\right\}$. Note that $|C_i|$ and $|C_i'|$ denote $c_i^o$ and $c_i'^o$, respectively. Since $s_i^l$ is a new element to the both of $C$ and $C'$, union operations between $\{s_i^l\}$ and each of $C$ and $C'$ increases the cardinality of $C$ and $C'$ by 1, respectively. In other words, $|C_i \cup \{s_i^l\}| = c_i^o + 1$, and $|C_i' \cup \{s_i^l\}| = c_i'^o + 1$.

Using (C.2), we can rewrite (C.1) as

$$\beta \cdot \left( (Q(C_i, u))^+ - \left( Q(C_i \cup \{s_i^l\}, u) \right)^+ \right) \leq \beta \cdot \left( (Q(C_i', u))^+ - \left( Q(C_i' \cup \{s_i^l\}, u) \right)^+ \right). \tag{C.3}$$

Depending on the sign of $Q(C_i, u)$, $(Q(C_i, u))^+$ can be evaluated as either $Q(C_i, u)$ or 0. Accordingly, we divide the cases for different signs of $Q(C_i, u)$ and prove that (C.3) holds for every case.

On the other hand, as $Q(C_i, u)$ is a decreasing function of $|C \cap S_i|$, $Q(C_i, u), Q(C_i', u), Q(C_i \cup \{s_i^l\}, u), Q(C_i' \cup \{s_i^l\}, u)$ can be ordered according to their values. We can utilize this order to exclude invalid cases.

From the decreasing property of $Q(C_i, u)$,

$$Q(C_i, u) \geq Q(C_i \cup \{s_i^l\}, u). \tag{C.4}$$

Furthermore, $C_i' \subset C_i$ implies

$$Q(C_i, u) \leq Q(C_i', u). \tag{C.5}$$

Based on (C.4) and (C.5), the followings are the valid cases for different signs of $Q(C_i, u)$.

1) $Q(C_i \cup \{s_i^l\}, u) > 0$
2) $Q(C_i', u) < 0$
3) $Q(C_i, u) > 0, Q(C_i \cup \{s_i^l\}, s) < 0, Q(C_i', s) > 0, Q(C_i' \cup \{s_i^l\}, u) > 0$
4) $Q(C_i, u) < 0, Q(C_i \cup \{s_i^l\}, u) < 0, Q(C_i', u) > 0, Q(C_i' \cup \{s_i^l\}, u) > 0$
5) $Q(C_i, u) > 0, Q(C_i \cup \{s_i^l\}, u) < 0, Q(C_i', u) > 0, Q(C_i' \cup \{s_i^l\}, u) < 0$
6) $Q(C_i, u) < 0, Q(C_i \cup \{s_i^l\}, u) < 0, Q(C_i', u) > 0, Q(C_i' \cup \{s_i^l\}, u) < 0$

The proofs for Cases 3-6 can be readily extended from the proof of Cases 1-2. Hence, we provide the proofs for Case 1-2 and omit the proofs for Case 3-6 due to space limitation.

*1) Case 1 ( $Q(C_i \cup \{s_i^l\}, u) > 0$):* Since $Q(C_i \cup \{s_i^l\}, u)$ is the minimum among $Q(C_i, u), Q(C_i', u), Q(C_i \cup \{s_i^l\}, u), Q(C_i' \cup \{s_i^l\}, u)$ for given $u$, all the terms of the data load are positive in this case. Thus, the increment of $G(C_i)$ can be represented as follows.

$$G\left(C_i \cup \{s_i^l\}\right) - G(C_i) = \beta \cdot \left(Q(C_i, u) - Q(C_i \cup \{s_i^l\}, u)\right), \tag{C.6}$$

$$= \beta. \tag{C.7}$$

Similarly, the increment of $G(C_i')$ becomes

$$G\left(C_i' \cup \{s_i^l\}\right) - G(C_i') = \beta \cdot \left(Q(C_i', u) - Q(C_i' \cup \{s_i^l\}, u)\right), \tag{C.8}$$

$$= \beta. \tag{C.9}$$

Consequently, for $Q(C_i \cup \{s_i^l\}, u) > 0$, (C.1) holds.

*2) Case 2 ($Q(C_i', u) < 0$):* For this case, all the terms are less than or equal to zero since $Q(C_i', u)$, the maximum value, is less than zero. Hence, we can rewrite the increments of $G(C_i)$ and $G(C_i')$ as follows.

$$G\left(C_i \cup \{s_i^l\}\right) - G(C_i) = \beta \cdot \left(\left(Q(C_i \cup \{s_i^l\}, u)\right)^+ - (Q(C_i, u))^+\right), \tag{C.10}$$

$$= 0. \tag{C.11}$$

$$G\left(C_i' \cup \{s_i^l\}\right) - G(C') = \beta \cdot \left(\left(Q(C_i' \cup \{s_i^l\}, u)\right)^+ - \left(Q(C_i', u)\right)^+\right), \tag{C.12}$$

$$= 0. \tag{C.13}$$

As a result, for $Q(C_i', u) < 0$, (C.1) holds.

For all possible cases, $G(C_i)$ shows submodularity. This completes the proof.

# Appendix D

## Proof of the Lemma 2

Since $L - c_i^o \geq 0$ for any $i$, $g(u) \geq -u \left\lfloor \frac{L}{u\mu} \mathbb{P}\left[\rho > \tau | u = 1\right] \log(1+\tau) \right\rfloor$. By the definition of flooring function, $u \left\lfloor \frac{L}{u\mu} \mathbb{P}\left[\rho > \tau | u = 1\right] \log(1+\tau) \right\rfloor \leq \frac{L}{\mu} \mathbb{P}\left[\rho > \tau | u = 1\right] \log(1+\tau)$. Therefore,

$$g(u) \geq -\frac{L}{\mu} \mathbb{P}\left[\rho > \tau | u = 1\right] \log(1+\tau), \tag{D.1}$$

$$= T \cdot c, \tag{D.2}$$

where $c = -L \cdot \mathbb{P}\left[\rho > \tau | u = 1\right] \log(1+\tau)$.

# Appendix E

## Proof of the Proposition 3

Since $(g(u))^+ \geq g(u)$ for any $u$,

$$\left| \overline{D(c_i^o)} - \sum_{i=1}^{F} f_i \cdot \left( \sum_{n=0}^{\infty} p_n \sum_{\left\{d_i^1, \cdots, d_i^n\right\} \in \mathcal{E}} p_{\left\{d_i^1, \cdots, d_i^n\right\}} g(u) \right)^+ \right| \leq \left| \overline{D(c_i^o)} - \sum_{i=1}^{F} f_i \sum_{n=0}^{\infty} p_n \sum_{\left\{d_i^1, \cdots, d_i^n\right\} \in \mathcal{E}} p_{\left\{d_i^1, \cdots, d_i^n\right\}} g(u) \right|, \tag{E.1}$$

$$\leq \left| \sum_{i=1}^{F} f_i \sum_{n=0}^{\infty} p_n \sum_{\left\{d_i^1, \cdots, d_i^n\right\} \in \mathcal{E}} p_{\left\{d_i^1, \cdots, d_i^n\right\}} g(u) \right|, \tag{E.2}$$

$$\leq T \cdot |c|, \tag{E.3}$$

where $c = -L \cdot \mathbb{P}\left[\rho > \tau | u = 1\right] \log(1+\tau)$.

# Appendix F

## Proof of the Proposition 4

Suppose $\left\{c_{i,\text{opt}}^o\right\}$ does not satisfy (22). In other words, $\exists k \in \mathcal{F}$ such that

$$c_{k,\text{opt}}^o > L - \nu. \tag{F.1}$$

Then, we can consider an another content placement $\left\{c_i^{o*}\right\}$ which is defined as

$$c_i^{o*} = \begin{cases} L - \nu & \text{for } i = k \\ c_{m,\text{opt}}^o + c_{k,\text{opt}}^o - \nu & \text{for } i = m \\ c_{i,\text{opt}}^o & \text{for } i \neq k \text{ and } i \neq m \end{cases}, \tag{F.2}$$

where $m$ is an index which satisfies

$$c_{m,\text{opt}}^o + c_{k,\text{opt}}^o < 2\nu. \tag{F.3}$$

In other words, $\left\{c_i^{o*}\right\}$ is made from $\left\{c_{i,\text{opt}}^o\right\}$ by truncating cached content by $\nu$ and using the truncated portion to cache another content more.

Then, if we compare the data load of a BS for two content placements $\left\{c_i^{o*}\right\}$ and $\left\{c_{i,\text{opt}}^o\right\}$,

$$\Delta = \sum_{i=1}^{F} f_i \cdot \left(L - c_i^{o*} - \nu\right)^+ - \sum_{i=1}^{F} f_i \cdot \left(L - c_{i,\text{opt}}^o - \nu\right)^+, \tag{F.4}$$

$$= f_m \cdot \left[\left(L - c_m^{o*} - \nu\right)^+ - \left(L - c_{m,\text{opt}}^o - \nu\right)^+\right]. \tag{F.5}$$

From the definition of $\left\{c_i^{o*}\right\}$ (F.2), it is obvious that $c_m^{o*} > c_{m,\text{opt}}^o$. Hence,

$$\left(L - c_m^{o*} - \nu\right)^+ \leq \left(L - c_{m,\text{opt}}^o - \nu\right)^+. \tag{F.6}$$

As a result, $\Delta \leq 0$ which contradicts to the optimality of $\left\{c_{i,\text{opt}}^o\right\}$. This completes the proof.

## Appendix G

## Proof of the Theorem 1

By Proposition 4, we can replace the constraint (20), without loss of optimality, by the following inequality

$$0 \leq c_i^o \leq L - \nu. \tag{G.1}$$

Hence, **P2** is equivalent to

$$\mathbf{P2}' : \min_{c_i} \sum_{i=1}^{F} f_i \cdot \left(L - c_i^o - \nu\right)^+ \tag{G.2}$$

$$\text{s.t. } 0 \le c_i^o \le L - \nu, \ \forall i \in \mathcal{F}, \tag{G.3}$$

$$\sum_{i=1}^{F} c_i \le M. \tag{G.4}$$

For $0 \le c_i^o \le L - \nu$, $\left(L - c_i^o - \nu\right)^+ = L - c_i^o - \nu$. Thus, the objective of **P2′** becomes

$$\sum_{i=1}^{F} f_i \cdot \left(L - c_i^o - \nu\right)^+ = \sum_{i=1}^{F} f_i \cdot \left(L - c_i^o - \nu\right). \tag{G.5}$$

If we omit constant terms irrelevant to $c_i^o$ and change minimization into maximization by removing the negative sign of $c_i^o$, **P2′** becomes equivalent to

$$\mathbf{P2}'' : \max_{c_i} \sum_{i=1}^{F} f_i c_i^o \tag{G.6}$$

$$\text{s.t. } 0 \le c_i^o \le L - \nu, \ \forall i \in \mathcal{F}, \tag{G.7}$$

$$\sum_{i=1}^{F} c_i^o \le M. \tag{G.8}$$

Suppose $\left\{c_{i,\mathrm{OM}}^o\right\}$ given in (24) is not optimal for **P2″**, there exists an optimal solution $\left\{c_i^{o*}\right\}$ such that

$$\sum_{i=1}^{F} f_i c_i^{o*} > \sum_{i=1}^{F} f_i c_{i,\mathrm{OM}}^o. \tag{G.9}$$

Then,

$$\Delta = \sum_{i=1}^{F} f_i c_{i,\mathrm{OM}}^o - \sum_{i=1}^{F} f_i c_i^{o,*}, \tag{G.10}$$

$$= \sum_{i \in \mathcal{D}_k^-} f_i \cdot \left(c_{i,\mathrm{OM}}^o - c_i^{o,*}\right) + \sum_{j \in \mathcal{D}_k^+} f_j \cdot \left(c_{j,\mathrm{OM}}^o - c_j^{o,*}\right), \tag{G.11}$$

where $\mathcal{D}_k^-$ and $\mathcal{D}_k^+$ are defined as $\mathcal{D}_k^- = \left\{i \le k | c_{i,\mathrm{OM}}^o \ne c_i^{o,*}\right\}$ and $\mathcal{D}_k^+ = \left\{j > k | c_{j,\mathrm{OM}}^o \ne c_j^{o,*}\right\}$, respectively. Moreover, because of feasibility of $\left\{c_i^{o,*}\right\}$, $\left\{c_i^{o,*}\right\}$ must satisfy the constraint in

(G.7). Consequently, $c_i^{o*} \leq L - \nu$. Also, from the definition of $\left\{c_{i,\mathrm{OM}}^o\right\}$ given in (24), if $i < k$, $c_{i,\mathrm{OM}}^o = L - \nu$. Hence, for $i < k$,

$$c_{i,\mathrm{OM}}^o - c_i^{o,*} \geq 0. \tag{G.12}$$

From (G.12), every term in the first summation of (G.11) is greater than or equal to 0. Furthermore, since $f_k$ is the minimum among $\{f_1, f_2, \cdots, f_k\}$, we can make the following lower bound.

$$\Delta = \sum_{i \in \mathcal{D}_k^-} f_i \cdot \left(c_{i,\mathrm{OM}}^o - c_i^{o,*}\right) + \sum_{j \in \mathcal{D}_k^+} f_j \cdot \left(c_{j,\mathrm{OM}}^o - c_j^{o,*}\right), \tag{G.13}$$

$$\geq f_k \sum_{i \in \mathcal{D}_k^-} \left(c_{i,\mathrm{OM}}^o - c_i^{o,*}\right) + \sum_{j \in \mathcal{D}_k^+} f_j \cdot \left(c_{j,\mathrm{OM}}^o - c_j^{o,*}\right). \tag{G.14}$$

From the definition of $\left\{c_{i,\mathrm{OM}}^o\right\}$, for $j > k$, $c_{j,\mathrm{OM}}^o = 0$. Hence, the lower bound becomes

$$f_k \sum_{i \in \mathcal{D}_k^-} \left(c_{i,\mathrm{OM}}^o - c_i^{o,*}\right) + \sum_{j \in \mathcal{D}_k^+} f_j \cdot \left(c_{j,\mathrm{OM}}^o - c_j^{o,*}\right) = f_k \sum_{i \in \mathcal{D}_k^-} \left(c_{i,\mathrm{OM}}^o - c_i^{o,*}\right) - \sum_{j \in \mathcal{D}_k^+} f_j c_j^{o,*}. \tag{G.15}$$

For Zipf distribution, since $f_k \geq f_j$ for $j > k$,

$$f_k \sum_{i \in \mathcal{D}_k^-} \left(c_{i,\mathrm{OM}}^o - c_i^{o,*}\right) - \sum_{j \in \mathcal{D}_k^+} f_j c_j^{o,*} \geq f_k \sum_{i \in \mathcal{D}_k^-} \left(c_{i,\mathrm{OM}}^o - c_i^{o,*}\right) - f_k \sum_{j \in \mathcal{D}_k^+} c_j^{o,*}, \tag{G.16}$$

$$= f_k \cdot \left(\sum_{i \in \mathcal{D}_k^-} c_{i,\mathrm{OM}}^o - \sum_{j \in \mathcal{D}} c_j^{o,*}\right). \tag{G.17}$$

Definitely, any optimal content placements should use all available memory, $\sum_{i \in \mathcal{D}_k^-} c_{i,\mathrm{OM}}^o = \sum_{i \in \mathcal{D}} c_i^{o,*}$. Therefore, $f_k \cdot \left(\sum_{i \in \mathcal{D}_k^-} c_{i,\mathrm{OM}}^o - \sum_{j \in \mathcal{D}} c_j^{o,*}\right) = 0$.

Finally, we have $\Delta \geq 0$. As a consequence,

$$\sum_{i=1}^{F} f_i c_i^{o,*} \leq \sum_{i=1}^{F} f_i c_{i,\mathrm{OM}}^o. \tag{G.18}$$

However, (G.18) contradicts to (G.9). By contradiction, $\left\{c_{i,\mathrm{opt}}^o\right\}$ is optimal.

## Appendix H

## Proof of the Lemma 3

Since $L - c_i^o \geq 0$ for any $i$, $g'(u) \geq -u \left\lfloor \frac{L}{\mu}\mathbb{P}\left[\rho > \tau|u\right]\log(1+\tau)\right\rfloor$. As the flooring function is less than or equal to the argument, $g'(u) \geq -u\frac{L}{\mu}\mathbb{P}\left[\rho > \tau|u=1\right]\log(1+\tau)$.

Define $u\left(\int_0^R \frac{x^\alpha}{x^\alpha+\tau r^\alpha}\frac{2x}{R^2}dx\right)^{u-1}$ as $A(u)$. Then,

$$u\frac{L}{\mu}\mathbb{P}\left[\rho > \tau|u\right]\log(1+\tau) = \frac{L}{\mu}\log(1+\tau)\int_0^R \exp\left(-r^\alpha\left(\frac{p}{\sigma^2}\right)^{-1}\tau\right)A\left(u\right)\frac{2r}{R^2}dr. \tag{H.1}$$

It can be easily shown that $A(u)$ has a single maximum point at $u = -\left(\ln\left(\int_0^R \frac{x^\alpha}{x^\alpha+\tau r^\alpha}\frac{2x}{R^2}dx\right)\right)^{-1}$. Hence, for any $u$, $A(u) \leq A\left(-\left(\ln\left(\int_0^R \frac{x^\alpha}{x^\alpha+\tau r^\alpha}\frac{2x}{R^2}dx\right)\right)^{-1}\right)$. As a result,

$$\begin{aligned} &u\frac{L}{\mu}\mathbb{P}\left[\rho > \tau|u\right]\log(1+\tau) \\ &\leq \frac{L}{\mu}\log(1+\tau)\int_0^R \exp\left(-r^\alpha\left(\frac{p}{\sigma^2}\right)^{-1}\tau\right)A\left(-\left(\ln\left(\int_0^R \frac{x^\alpha}{x^\alpha+\tau r^\alpha}\frac{2x}{R^2}dx\right)\right)^{-1}\right)\frac{2r}{R^2}dr. \end{aligned} \tag{H.2}$$

Since $T = \frac{1}{\mu}$, if we set $c'$ as

$$c' = -L\log(1+\tau)\int_0^R \exp\left(-r^\alpha\left(\frac{p}{\sigma^2}\right)^{-1}\tau\right)A\left(-\left(\ln\left(\int_0^R \frac{x^\alpha}{x^\alpha+\tau r^\alpha}\frac{2x}{R^2}dx\right)\right)^{-1}\right)\frac{2r}{R^2}dr, \tag{H.3}$$

we can obtain $g'(u) \geq T \cdot c'$.